\documentclass[epj]{svjour}

\newcommand{\ststbar}{\ensuremath{\tilde{t}_{1}\bar{\tilde{t}}_{1}}}
\newcommand{\stopone}{\ensuremath{\tilde{t}_{1}}}
\newcommand{\chargino}{\ensuremath{\tilde{\chi}^{\pm}_{1}}}
\newcommand{\charginoplus}{\ensuremath{\tilde{\chi}^{+}_{1}}}
\newcommand{\neutralino}{\ensuremath{\tilde{\chi}^{0}_{1}}}
\newcommand{\ttbar}{\ensuremath{t\bar{t}}}
\newcommand{\ppbar}{\ensuremath{p\bar{p}}}
\newcommand{\Dzero}{D{\O}}
\newcommand{\pt}{\ensuremath{p_{T}}}
\newcommand{\met}{\ensuremath{\not\!\!{E}_{T}}}
\newcommand{\prospino}{\textsc{Prospino}}

\usepackage{graphicx}
\usepackage{fancyhdr}
\usepackage{url}
\usepackage{subfigure}
\def\makeheadbox{{
 }}
\def\mytitle{My title}
\def\myauthors{My name}
\def\mytype{My type of session}
\def\mysession{My session}

\def\mytitle{Search for Scalar Top Admixture in the \ttbar\ Lepton+Jets 
Channel at \Dzero} 
\def\myauthors{Su-Jung Park}    
\def\mytype{Contributed Talk}
\def\mysession{Colliders - SUSY Phenomenology}

\begin{document}
\title{Search for Scalar Top Admixture in the \ttbar\ Lepton+Jets Channel at \Dzero}
\subtitle{}
\author{Su-Jung Park 
on behalf of the \Dzero\ Collaboration
}                     
%
%
\institute{University of Rochester,
Department of Physics\&Astronomy,
Bausch\&Lomb Hall,
P.O. Box 270171,
Rochester, NY 14627-0171, USA}
%
\date{}
\abstract{
A search has been performed for scalar top quark pair production 
in the lepton+jets channel in $\approx$ 1 fb$^{-1}$ of data. Kinematic 
differences between the exotic \ststbar\ and the dominant \ttbar\ 
process are used to separate the two possible contributions. For scalar
top quark masses of 145--175~GeV and chargino masses of 
105--135~GeV we obtain upper cross section limits at 95\% confidence 
level for \ststbar\ production that are a factor of $\approx$7--12 
higher than expected for the Minimal Supersymmetric Standard Model (MSSM).
} 
\maketitle
\section{Introduction} \label{sec_intro}
\subsection{Scalar Top Production and Decay} \label{sec_introstop}
Scalar top (stop) quarks are mainly produced in pairs with essentially
the same diagrams as top pairs. The theoretical cross section at a
center of mass energy of 1.96~TeV for a stop quark of the mass 175~GeV
is 0.579~pb \cite{prospino}. As comparison, for a top quark of the same 
mass, the cross section is 6.77~pb \cite{topxsec}. The stop quark pair 
production cross section highly depends on the mass of the stop quark.

Even with the assumption that R-parity is conserved there are many
possible decays for the stop quark. The branching ratios depend on the
SUSY parameters chosen, in particular the masses of the supersymmetric
particles involved. The decays $\stopone \to c \neutralino$ \cite{d0stoplimit1}
and $\stopone \to b \ell^+ \tilde{\nu}_\ell$ \cite{d0stoplimit2} have
already been explored at \Dzero\ in Run II. Another important decay channel
is the $\stopone \to b \charginoplus$ channel considered in this analysis
\cite{stopconfnote}.
We consider six mass points, for which we vary the stop quark mass from 
145 to 175~GeV and the chargino mass from 105 to 135~GeV, while keeping
the neutralino mass fixed to 50~GeV.

\subsection{Tevatron Collider and \Dzero\ Detector}
The Tevatron collider is located at Fermi National Accelerator Laboratory
(Fermilab) in Batavia, Illinois, USA. It collides protons and antiprotons
at a center of mass energy of 1.96~TeV. As of July 2007 it has delivered
3.2~fb$^{-1}$, of which \Dzero\ recorded 2.72~fb$^{-1}$. The analysis
presented uses approximately 900~pb$^{-1}$. 

\Dzero\ is one of the two multi-purpose experiments at Fermilab \cite{D0RUNII}. 
Going 
from the inside out the main parts of the detector consist of a tracking 
system inside a magnetic field, a calorimeter, and a muon system. The
detector allows for the reconstruction of tracks, vertices, electrons,
photons, jets, missing transverse energy and muons to high pseudorapitidy
regions.

\section{Search for Scalar Top} \label{sec_search}

\subsection{Event Signature}
As reported in Section~\ref{sec_introstop} each of the produced stop quarks
decays to a chargino \chargino\ and a $b$-quark. The chargino then decays to 
a $W$~boson and a neutralino \neutralino. The subsequent decay of the 
$W$~boson determines the event topology just as in top quark pair production 
(\ttbar) events. The resulting \ststbar\ event signature is consequently very 
similar to the \ttbar\ signature, thus making it possible for the \ststbar\ 
signal to be contained in the \ttbar\ event sample. The only difference to
top pair production are the additional neutralinos in the event.
For this analysis, we 
consider the decay channel with one $W$~boson decaying to hadrons and the 
other one to leptons. It does not play a role whether the $W$~boson is 
on-shell or off-shell, both scenarios produce the same signature.
The resulting final state consists of one high-\pt\ lepton, missing transverse 
energy \met\ from the neutrino and the neutralinos, two $b$-jets, and two 
light quark jets. This will be referred to as lepton+jets channel. The 
analysis is performed separately in the electron+jets ($e$+jets) channel,
where the lepton is an electron (including one from $\tau$ decay) and the 
muon+jets ($\mu$+jets) channel, where the lepton is a muon (also including 
one from $\tau$ decay). The result of the two channels is then combined.

\subsection{Background Processes}
\begin{figure}[h!tbp]
	\begin{center}
	\includegraphics*[width=0.45\textwidth]{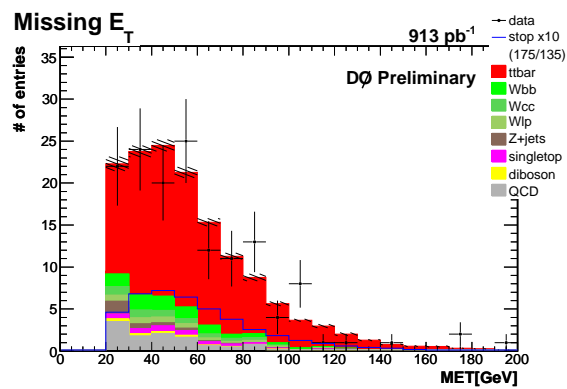}
	\end{center}
	\caption{\label{fig_controlplotmet} Data-MC agreement
	for missing transverse energy \met\ in the $e$+jets channel
	after the preselection with $\ge$4 jets and $b$-tagging. The
	blue line shows the distribution for the 175/135 signal point
	ten times enhanced.}
\end{figure}

Because of their similarity to the signal \ttbar\ events dominate the
background and are very difficult to separate. The remaining background
processes are common to both \ststbar\ and \ttbar\ events, of which
QCD multijet production and $W$+jets dominate. Smaller contributions stem
from $Z$+jets production, single top production and diboson production.
For the common background processes the same methods as in \Dzero\ \ttbar\ 
analyses are used to reduce and estimate them. The preselection preferably
selects events containing real leptons, which reduces the QCD multijet
background. The difference between the data with a looser lepton quality
requirement and a tighter lepton quality requirement is used to estimate the
remaining QCD multijet background. Also using data the $W$+jets is estimated
before a requirement on the identification of a jet originating from a 
$b$-quark ($b$-jet) removes most of that background. The contribution
of the smaller background processes is calculated from their theoretical
cross sections. The exact treatment of the \ttbar\ background is discussed
in the next section, but its contribution is also derived from its
theoretical cross section. Table~\ref{tab_expsig} shows the number of 
expected signal events after the complete preselection for all six mass
points, Table~\ref{tab_expbg} shows the number of expected background events.
Figure~\ref{fig_controlplotmet} shows the data-MC agreement for one variable,
the missing transverse energy \met, after the preselection with $\ge$4 jets
and after $b$-tagging. The dominance of the \ttbar\ background is clearly
visible.

\begin{table}[h!tbp]
	\begin{center}
		\caption{\label{tab_expsig} Expected number of signal events
		after complete preselection for 913~pb$^{-1}$ in the electron+jets
		channel and 871~pb$^{-1}$ in the muon+jets channel.}
		\begin{tabular}{lrr}
        	\hline 
		$m_{\stopone}/m_{\chargino}$ & $e$+jets & $\mu$+jets \\
        	\hline 
		175/135 & 4.0 & 3.1 \\
		175/120 & 3.1 & 2.3 \\
		175/105 & 2.8 & 2.0 \\
		160/120 & 3.6 & 2.4 \\
		160/105 & 3.8 & 2.7 \\
		145/105 & 4.5 & 3.0 \\
         	\hline 
        	\end{tabular}
      	\end{center}
\end{table}

\begin{table}[h!tbp]
	\begin{center}
		\caption{\label{tab_expbg} Expected number of background events
		after complete preselection for 913~pb$^{-1}$ in the electron+jets
		channel and 871~pb$^{-1}$ in the muon+jets channel.}
		\begin{tabular}{lrr}
        	\hline 
		Background & $e$+jets & $\mu$+jets \\
        	\hline 
		\ttbar\    & 103.0 &  84.2 \\
		$Wbb$      &   8.5 &  11.1 \\
		$Wcc$      &   4.8 &   6.5 \\
		$W$light   &   3.8 &   4.0 \\
		$Z$+jets   &   2.8 &   3.3 \\
		single top &   3.1 &   2.5 \\
		diboson    &   1.4 &   1.2 \\
		multijet   &  10.7 &   3.2 \\
		\hline
		SUM        & 138.1 & 116.0 \\
		\hline
		Data       & 133   & 135 \\
         	\hline 
        	\end{tabular}
      	\end{center}
\end{table}

\subsection{Limit Setting Procedure}
\begin{figure}[h!tbp]
	\begin{center}
	\includegraphics*[width=0.45\textwidth]{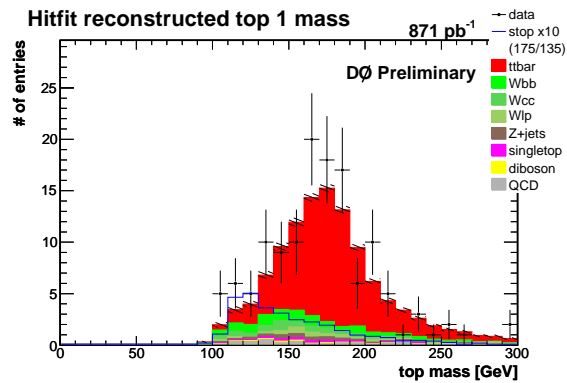}
	\end{center}
	\caption{\label{fig_hftopmass} Data-MC agreement
	for the reconstructed top mass in the $\mu$+jets channel
	after the preselection with $\ge$4 jets and $b$-tagging. The
	blue line shows the distribution for the 175/135 signal point
	ten times enhanced.}
\end{figure}

\begin{figure}[h!tbp]
	\begin{center}
	\includegraphics*[width=0.45\textwidth]{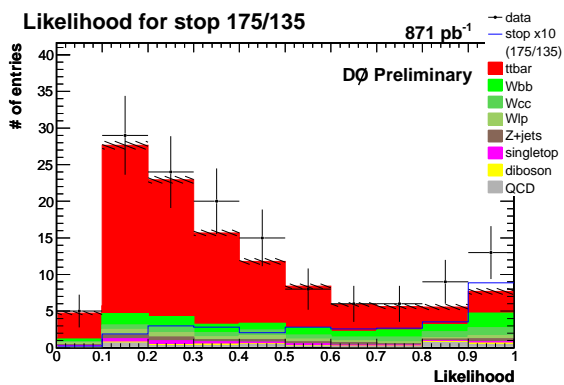}
	\end{center}
	\caption{\label{fig_lhood} Data-MC agreement
	likelihood discriminant in the $e$+jets channel
	after the preselection with $\ge$4 jets and $b$-tagging. The
	blue line shows the distribution for the 175/135 signal point
	ten times enhanced.}
\end{figure}

Although we have already estimated the \ttbar\ background using its
theoretical cross section, we still need a discrimination variable to
be able to separate it from the \ststbar\ signal. Because of the additional
neutralinos one might expect additional missing transverse energy \met, but 
since the neutralinos tend to be back-to-back, this is not the case and 
\met\ looks the same in stop and top events. There are, however, minor
differences in some distributions. Especially helpful is a kinematic fitter,
which reconstructs events to a \ttbar\ hypothesis. As an example, 
Figure~\ref{fig_hftopmass} shows the top mass as reconstructed by the 
kinematic fitter in the $\mu$+jets channel. Comparing the \ttbar\ in red to 
the blue line, which represents the signal, it can be seen that the 
distribution is very different for stop and top events. Combined into a
likelihood discriminant the separation is even more pronounced as seen in
Figure~\ref{fig_lhood}.

We use a Bayesian approach to extract limits from the likelihood discriminant
distributions \cite{limitsetting}. Under the assumption of a Poisson
distribution for observed counts a binned probability function is formed. 
It is a product over all bins of the likelihood discriminant distribution 
and can be extended to combine the $e$+jets and $\mu$+jets channels. For the
signal cross section we assume a flat nonnegative prior probability. By
integrating over the signal acceptances, background yields and integrated
luminosity with Gaussian priors for each systematic uncertainty we obtain the
posterior probability density as a function of the signal cross section. The
limit at 95\% confidence level is the point where the integral over the
posterior probability density reaches 95\% of its total. 

The expected results are derived on the sum of all preselected background 
Monte Carlo samples without a \ststbar\ contribution, but including the 
\ttbar\ contribution according to its theoretical value of 6.77~pb. They 
are shown together with the observed limits and the theoretically predicted
cross section in Table~\ref{tab_lim} for the $e$+jets and $\mu$+jets
channels combined.

\section{Result and Conclusion} \label{sec_result}
\begin{figure}[h!tbp]
	\begin{center}
	\includegraphics*[bb=380 0 557 173, clip, width=0.4\textwidth]{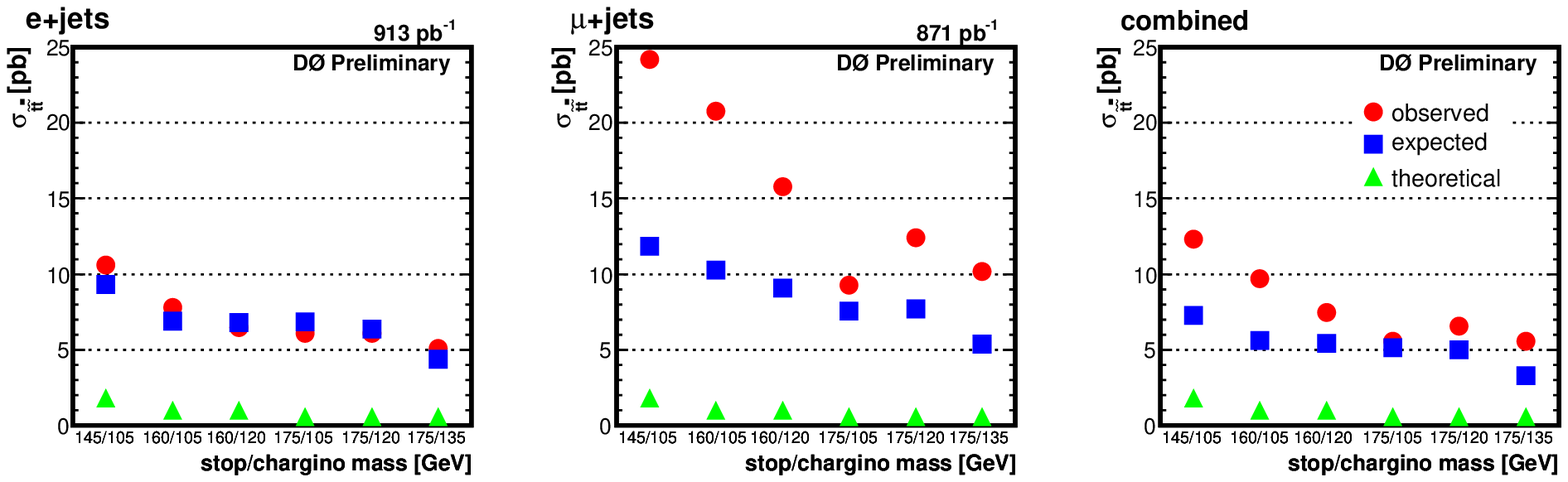}
	\end{center}
	\caption{\label{fig_limit1D} Observed and expected Bayesian limits at
	95\% confidence level on the \ststbar\ cross section and theoretical 
	cross section for \ststbar\ at each mass point for combined channels.}
\end{figure}

Table~\ref{tab_lim} shows the result for each mass point for $e$+jets 
and $\mu$+jets channels combined. The first three columns show the masses
of the stop quark, the chargino, and the neutralino. The fourth column
gives the theoretically predicted cross section for \ststbar\ production,
the fifth column shows the expected and the sixth column the observed
Bayesian limit at 95\% confidence level on the \ststbar\ cross section.
The result is also illustrated in Figure~\ref{fig_limit1D}. 
At this point we cannot exclude any of the stop masses, all  observed limits 
are a factor of $\approx$7-12 above the theoretical predictions.

\begin{table}[h!tbp]
	\begin{center}
		\caption{\label{tab_lim} 
		SUSY particle masses in GeV, the theoretical \ststbar\ cross 
		section, and the expected and observed Bayesian limits at 95\% 
		confidence level on the \ststbar\ cross section in pb for 
		combined channels.}
		\begin{tabular}{cccccc}
        	\hline 
		\multicolumn{3}{c}{SUSY masses} & theo & exp & obs \\
		$m_{\stopone}$ & $m_{\chargino}$ & $m_{\neutralino}$ & $\sigma_{\ststbar}$ & $\sigma_{\ststbar}$ & $\sigma_{\ststbar}$\\
		$[$GeV] & [GeV] & [GeV] & [pb] & [pb] & [pb]\\
        	\hline 
		175 & 135 & 50 & 0.579 & 3.28 & 5.57 \\
		175 & 120 & 50 & 0.579 & 4.97 & 6.58 \\
		175 & 105 & 50 & 0.579 & 5.16 & 5.55 \\
		160 & 120 & 50 & 1.00  & 5.42 & 7.45 \\
		160 & 105 & 50 & 1.00  & 5.63 & 9.71 \\
		145 & 105 & 50 & 1.80  & 7.27 & 12.32\\
         	\hline 
        	\end{tabular}
      	\end{center}
\end{table}

%
%

\end{document}